# Ionized Impurity and Surface Roughness Scattering Rates of Electrons in Semiconductor Structures with One-Dimensional Electron Gas and Broadened Energy Levels

D.V. Pozdnyakov[1], V.O. Galenchik, V.M. Borzdov,
F.F. Komarov, O.G. Zhevnyak

*Radiophysics and Electronics Department, Belarus State University,
Nezavisimosty av. 4, 220050 Minsk, Belarus*

**Abstract**

An approach to calculation of the ionized impurity and surface roughness scattering rates of electrons in very thin semiconductor quantum wires taking into account the energy level broadening is worked out. It is assumed that all the electrons in the structure are in the electric quantum limit. The screening is taken into account while considering the ionized impurity scattering. Comparison of the surface roughness scattering rates calculated using the exponential and Gaussian autocorrelation functions is done.

**Keywords**: *Ionized impurity; Surface roughness; Quantum wire*

## 1. Introduction

Currently the study of electrophysical properties of semiconductor structures with one-dimensional electron gas is of great interest [1–3] due to a number of unique quantum effects occurring in them. Those effects allow a new generation of semiconductor devices to be developed. One of the most efficient approaches to the study of electron transport in semiconductor structures is Monte Carlo simulation [4, 5]. The correct use of this method is only possible if a correct description of the scattering mechanisms is available. The scattering rates should be calculated with the maximum possible accuracy, i.e. with the secondary quantum effects taken into account [6]. In particular, one of such effects is the energy level broadening causing the energy uncertainty of electrons. The influence of this effect on the acoustic and polar optical phonon scattering rates is studied in Refs. [6–8]. Nevertheless, as far as we know the influence of this effect on the ionized impurity and surface roughness scattering rates has not yet been elucidated. So this study is devoted to the calculation of the ionized impurity and surface roughness scattering rates taking into account the energy level broadening.

## 2. Theory

Let us consider very thin quantum wires with rectangular cross-section when all the electrons are in the ground quantum state, i.e. so called electric quantum limit. In this case the energy level broadening is the same that the collisional broadening [6]. Moreover, only the backward scattering is considered since the forward elastic scattering (such as that caused by ionized impurity or surface roughness) does not change the electron energy and momentum. It should be noted that the perturbation of a quantum system caused by ionized impurities or surface roughness is constant in time and non-uniform in space. It differs from the phonon one which is uniform in space and varies with time [9]. In this connection it is necessary to present the electrons as the wave packets instead of monochromatic de Broglie waves when calculating the ionized impurity and surface roughness scattering rates [10]. Thus, the energy level broadening has a profound effect on the scattering matrix element as well as on the density of final states [7, 8, 11]. Then, using the results of Refs. [7–11], the following first order approximate formula for the transformation of the scattering matrix element can be obtained:

---

[1] pozdnyakov@bsu.by

$$|M_{br}(E \mp E)|^2 \approx |M_{id}(E \mp E + \Delta E)|^2, \qquad (1)$$

where $\Delta E$ is the energy level halfwidth [11], $E$ is the electron average kinetic energy, $M_{br}$ is the matrix element with the energy level broadening taken into account, and $M_{id}$ is the same with the energy level broadening neglected. The sign «–/+» corresponds to the forward/backward scattering.

The electron intrasubband scattering rate in the lowest level with the energy level broadening taken into account can be presented as [7, 8, 11]

$$W(E) = \frac{L_x \sqrt{2m_x}}{2\hbar^2} |M_{br}(E)|^2 \sqrt{\frac{\Delta E + \sqrt{\Delta E^2 + E^2}}{\Delta E^2 + E^2}}, \qquad (2)$$

where $L_x$ is the quantum wire length, $m_x$ is the effective electron mass in the direction along the quantum wire, and $\hbar$ is the Planck constant. Thus, according to Eqs. (1) and (2) the problem of calculation of the ionized impurity and surface roughness scattering rates is reduced to that of calculating the matrix elements $M_{id}^{ii}$ and $M_{id}^{sr}$, respectively.

The value of $|M_{id}^{ii}|^2$ can be obtained by the following formulae derived using the results of Refs. [12–14]:

$$|M_{id}^{ii}(k)|^2 = \left(\frac{Z_0 e^2}{2\pi\varepsilon}\right)^2 \frac{N_L}{L_x} \left((2k\rho)^{-2} + K_0^2(2k\rho) - K_1^2(2k\rho)\right), \qquad (3)$$

$$\rho = \sqrt{\frac{L_y L_z}{\pi}}, \qquad (4)$$

$$k = \frac{\sqrt{2m_x E}}{\hbar}, \qquad (5)$$

where $Z_0$ is the ion relative charge, $e$ is the electron charge, $\varepsilon$ is the semiconductor dielectric permittivity, $N_L$ is the linear concentration of ionized impurities in the quantum wire, $K_n$ is the modified Bessel function of the second kind and $n$-th order, $L_y$ and $L_z$ are the transverse dimensions of the quantum wire. Moreover, it should be noted that Eq. (3) was obtained under the following assumptions: (i) the Sakaki's approximation for electron wave function in the YZ-plane is used [12], (ii) the screening effect is neglected, (iii) the dimensions $L_y$ and $L_z$ should be of the same order ($L_y \sim L_z$). Then, using the results of Refs. [12–14] and taking into account the screening effect as well as the exact expression for electron wave function in the YZ-plane, the following formulae can be derived:

$$|M_{id}^{ii}(k)|^2 = \left(\frac{Z_0 e^2}{2\pi\varepsilon}\right)^2 \frac{N_L}{L_x} F^2(k), \qquad (6)$$

$$F(k) \approx \frac{1}{k\rho} \int_0^\infty \left(\sqrt{1+x^2} - x\right)\left(\alpha + \frac{1}{\sqrt{1+x^2}}\right) \exp(-\alpha x) \sin(2k\rho x) dx, \qquad (7)$$

$$\alpha = 2\rho N_L. \qquad (8)$$

Finally according to Eqs. (1) and (6) the ionized impurity scattering matrix element with the energy level broadening accounted for can be presented as

$$|M_{br}^{ii}(k)|^2 = \left(\frac{Z_0 e^2}{2\pi\varepsilon}\right)^2 \frac{N_L}{L_x} F^2\left(\sqrt{k^2 + \chi^2}\right), \qquad (9)$$

$$\chi = \frac{\sqrt{m_x \Delta E}}{\hbar}. \qquad (10)$$

Let us now consider the surface roughness scattering. According to Refs. [15–18] the follow-ing formulae for the scattering matrix element can be derived by using the exponential and Gaussian autocorrelation functions, respectively:

$$\left|M_{id}^{GS}(k)\right|^2 = \frac{\sqrt{\pi}\Lambda\Gamma^2}{L_x}\exp(-k^2\Lambda^2), \qquad (11)$$

$$\left|M_{id}^{EX}(k)\right|^2 = \frac{\sqrt{\pi}\Lambda\Gamma^2}{L_x}\frac{1}{1+\pi k^2\Lambda^2}, \qquad (12)$$

$$\Gamma^2 = 2\delta_y^2\left(\frac{\partial E_0}{\partial L_y}\right)^2 + 2\delta_z^2\left(\frac{\partial E_0}{\partial L_z}\right)^2, \qquad (13)$$

where $\Lambda$ is the roughness correlation length, $\delta_y$ and $\delta_z$ are the root-mean-square deviations of the rough quantum wire boundary from the plane which is normal to the axes $Z$ and $Y$, respectively, and $E_0$ is the energy level of the ground quantum state. Moreover, for the Gaussian autocorrelation function the standard equation is used [15–18]:

$$\langle\delta(x)\delta(x')\rangle_{GS} = \delta^2\exp\left(-\frac{(x-x')^2}{\Lambda^2}\right), \qquad (14)$$

and for the exponential one [18]

$$\langle\delta(x)\delta(x')\rangle_{EX} = \delta^2\exp\left(-\frac{|x-x'|}{\lambda}\right). \qquad (15)$$

To provide the equivalence between functions $\langle\delta(x)\delta(x')\rangle_{GS}$ and $\langle\delta(x)\delta(x')\rangle_{EX}$ the values of $\Lambda$ and $\lambda$ are chosen to satisfy the equation

$$\int_{-\infty}^{+\infty}\langle\delta(x)\delta(x')\rangle_{GS}\,dx = \int_{-\infty}^{+\infty}\langle\delta(x)\delta(x')\rangle_{EX}\,dx, \qquad (16)$$

which yields $\lambda = \Lambda\sqrt{\pi/4}$.

Let us compare the adequacy of the description of scattering processes when employing the exponential and Gaussian autocorrelation functions. To perform this comparison we are evaluating the surface roughness scattering rates at high electron kinetic energies by using a different non-matrix approach. At high kinetic energies the electron momentum uncertainty is much less than the momentum itself. At the same time, the electron can be localized in space with dimensions much less than $\Lambda$, and momentum uncertainty $\hbar\langle\Delta k\rangle$ at $\hbar k \to +\infty$ is much less than $\hbar k$. Then scattering acts are equivalent to reflection of electrons from the region where a sharp potential perturbation takes place. The scattering acts are independent of each other, i.e. non-coherent scattering takes place [19] due to the possibility of particle localization in the region with dimension less than $\Lambda$ [15]. Moreover, because of the high velocity of charge carriers the disturbed potential due to roughness can be considered as the Kronig-Penney potential whose period on the order of magnitude is $\Lambda$ [15]. Thus the surface roughness scattering rate at high electron kinetic energy can be evaluated as

$$W_{SR}(E) \approx \frac{vR}{\Lambda} = \frac{1}{\Lambda}\sqrt{\frac{2E}{m_x}}\left(\frac{1-\sqrt{1-\Gamma^2 E^{-2}}}{1+\sqrt{1-\Gamma^2 E^{-2}}}\right) \approx \frac{\Gamma^2}{2\Lambda E\sqrt{2m_x E}}, \qquad (17)$$

where $v$ is the electron group velocity, $R$ is the electron reflection coefficient from the point where a step of the Kronig-Penney potential takes place.

While comparing the Eqs. (2), (11), (12) and (17), the following conclusions can be made: (i) the scattering rate calculated by using the Gaussian autocorrelation function inadequately describes scattering at $E >> \hbar^2/(m_x\Lambda^2)$, (ii) the surface roughness scattering rate calculated by using the exponential autocorrelation function provides a correct description at various energies. Moreover, the results of calculations by using $\left|M_{id}^{EX}(k)\right|^2$ are close to those obtained by employ-ing $\left|M_{id}^{GS}(k)\right|^2$ at $E < \hbar^2/(m_x\Lambda^2)$. At $E > \hbar^2/(m_x\Lambda^2)$ the value of $\left|M_{id}^{EX}(k)\right|^2$ shows that the scattering rate as a function of energy is similar to that for the non-coherent scattering described by formula (17). It should be noted that the same conclusion is made in Ref. [18] from the comparison of results of Monte Carlo simulation with the experimental data.

Now the final equation for the surface roughness scattering matrix element taking into account Eqs. (1), (12) and the energy level broadening can be expressed as

$$\left|M_{br}^{sr}(k)\right|^2 = \frac{\sqrt{\pi}\Lambda\Gamma^2}{L_x}\frac{1}{1+\pi(k^2+\chi^2)\Lambda^2}. \tag{18}$$

**3. Results and discussion**

As an example, in Fig. 1 and 2 the ionized impurity and surface roughness backward scattering rates in GaAs quantum wire with infinite potential barriers at the boundaries are plotted against the kinetic energy at temperature $T = 300$ K, $L_y = L_z = 40$ Å, $\Lambda = 60$ Å, $\delta_y = \delta_z = 1.415$ Å and $N_L = 1.6\times10^6$ m$^{-1}$. The value of $\Delta E$ was chosen to be 7.5 meV [7, 8].

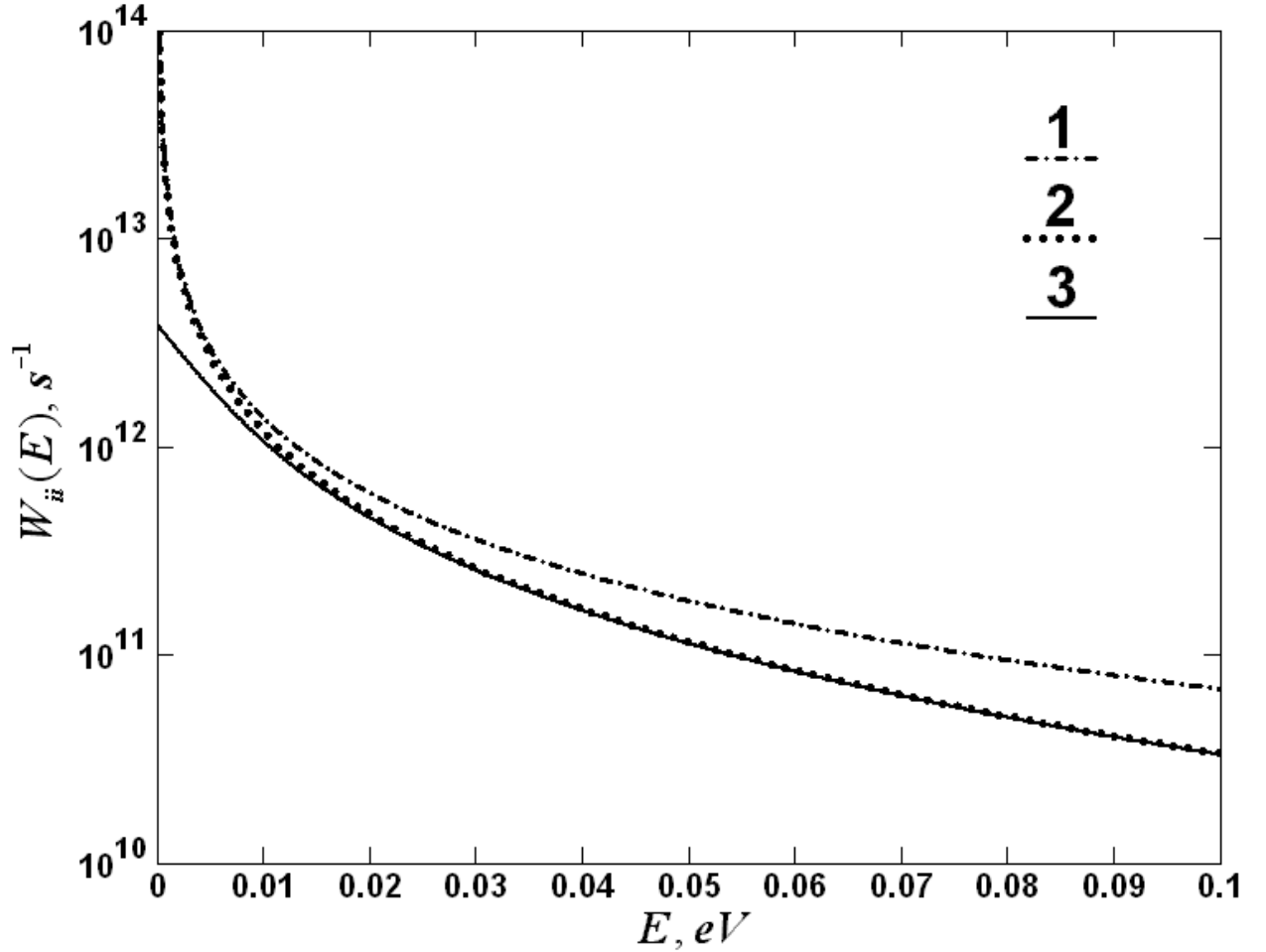

**Fig. 1. Ionized impurity backward scattering rates of electrons in GaAs quantum wire calculated by using:**
*curve 1 – the Sakaki's approximation for electron wave function in the YZ-plane neglecting both the screening effect and the energy level broadening,*
*curve 2 – the exact expression for electron wave function in the YZ-plane with the screening effect taken into account but neglecting the energy level broadening,*
*curve 3 – the exact expression for electron wave function in the YZ-plane with both the screening effect and the energy level broadening taken into account.*

Fig. 1 reveals that the ionized impurity scattering rate calculated by taking into account the screening effect and the exact expression for electron wave function is lower than the rate calculated in the Sakaki's approximation.

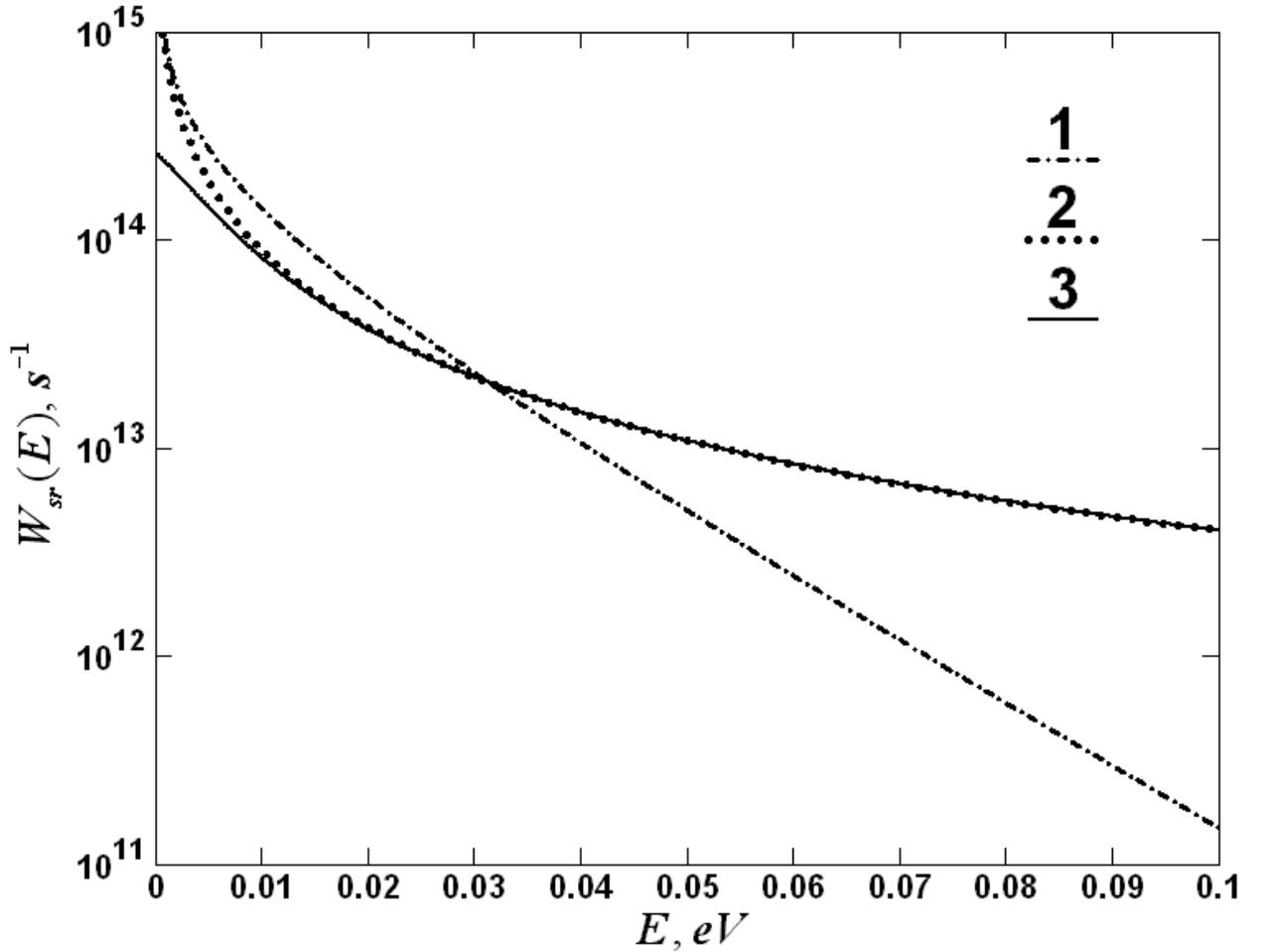

**Fig. 2.** Surface roughness backward scattering rates of electrons in GaAs quantum wire calculated by using:
*curve* 1 – *the Gaussian autocorrelation function with the energy level broadening neglected,*
*curve* 2 – *the exponential autocorrelation function with the energy level broadening neglected,*
*curve* 3 – *the exponential autocorrelation function with the energy level broadening taken into account.*

According to Fig. 2 the surface roughness scattering rate calculated with the exponential autocorrelation function is close to the rate obtained by employing the Gaussian autocorrelation function at low kinetic energies. At high energies the first one decreases less steeper than the second one. In both cases (the impurity and surface roughness scattering) accounting for the energy level broadening prevents the scattering rate from tending to infinity. A significant difference between the scattering rates calculated with the energy level broadening (i) taken into account and (ii) neglected is only observed at energies $E \in (0, \Delta E)$ what is in a good agreement with the results of Refs. [6–8].

Thus, the approaches to calculation of the ionized impurity and surface roughness scattering rates taking into account the energy level broadening are developed in this paper. The obtained results are in good agreement with known theoretical insights [12–18].